\documentclass[prl, aps, onecolumn, showpacs, amsmath, amssymb, amsfonts, floatfix, superscriptaddress, nobibnotes, 10pt]{revtex4}

\usepackage{calc}
\usepackage{latexsym}
\usepackage{float}
\usepackage{supertabular}
\usepackage{longtable}
\usepackage{graphicx, amsmath, amssymb, bm}

\usepackage{ifpdf}
\ifpdf
	\newcommand{\ignoreThis}[1]{}
	\usepackage{color}
	\definecolor{Gray}{rgb}{0.6,0,0}
\else
	\newcommand{\ignoreThis}[1]{#1}
	\usepackage[usenames,dvips]{color}
\fi
\usepackage[normalem]{ulem}

\newcommand{\be}{\begin{equation}}
\newcommand{\ee}{\end{equation}}
\newcommand{\bs}{\begin{split}}
\newcommand{\es}{\end{split}}

\begin{document}
\title{Genetic Toggle Switch in the Absence of Cooperative Binding: Exact Results}
\author{Tommaso Biancalani}
\affiliation{Department of Physics and Institute for Genomic
Biology, University of Illinois at Urbana-Champaign, Loomis Laboratory
of Physics, 1110 West Green Street, Urbana, Illinois 61801-3080, USA}
\author{Michael Assaf}
\affiliation{Racah Institute of Physics, Hebrew University of Jerusalem, Jerusalem 91904, Israel} %

\begin{abstract}
We present an analytical treatment of a genetic switch model consisting of two mutually inhibiting genes operating without cooperative binding of the corresponding transcription factors. Previous studies have numerically shown that these systems can exhibit bimodal dynamics  without possessing two stable fixed points at the deterministic level. We analytically show that bimodality is induced by the noise and find the critical repression strength that controls a transition between the bimodal and non-bimodal regimes. We also identify characteristic polynomial scaling laws of the mean switching time between bimodal states. These results, independent of the model under study, reveal essential differences between these systems and systems with cooperative binding, where there is no critical threshold for bimodality and the mean switching time scales exponentially with the system size.
\end{abstract}

\maketitle
Gene expression in living cells is regulated by transcription factors that bind to specific DNA sequences thereby promoting or repressing  transcription of genes. This mechanism allows for a ``digital'' response: when a cell has to make a decision, between expressing a certain protein, $A$, or another, $B$, a biochemical regulatory network leads the system to a state either dominated by $A$, or $B$. Such behavior is called bimodal. An example of such decision-making circuits is given by the genetic toggle switch in which two transcription factors mutually repress each other~\cite{ptashne1992phage,golding2011decision}. This and other genetic switches allow cells to switch between distinct phenotypic states and determine the cell's fate, in response to environmental stimuli and/or internal signals~\cite{elowitz2002stochastic, thattai2001intrinsic, hasty2000noise, mcadams1997stochastic}.

Genetic switches are found to exhibit distinct behaviors according to whether or not there is cooperative binding (CB) of transcription factors (see \textit{e.g.}~\cite{to2010noise} in the context of positive feedback). If CB is in play, more than a single transcription factor molecule can bind to the DNA sequence, and the binding probability depends on whether there are molecules already bound to the sequence. CB is a driver of bimodality and was previously thought to be a necessary condition for a bimodal behavior~\cite{gardner2000construction, warren2005chemical, allen2005sampling, schultz2008extinction}. This is since when CB is present in the rate equations, there are (at least) two stable fixed points corresponding to states rich in each type of transcription factors; in contrast, the absence of CB yields a single stable fixed point where the two transcription factors coexist.

Yet, in recent years, it has been shown in different models theoretically~\cite{samoilov2005stochastic,lipshtat2006genetic,loinger2007stochastic}, and experimentally~\cite{to2010noise}, that bimodality can emerge even without having bistability at the deterministic level. In~\cite{to2010noise}, bimodality has been reported in a synthetic budding yeast system, which concluded that the bimodal behavior is induced by demographic noise. In Refs.~\cite{lipshtat2006genetic, loinger2007stochastic}, the authors have numerically shown that a genetic toggle switch can exhibit a bimodal behavior due to demographic noise, even in the absence of CB. To this end, in Ref.~\cite{venegas2011analytical} the exclusive switch model (ESM) was analytically studied via the probability generating function. Yet, their analysis, valid only in limiting cases, cannot uncover how demographic noise gives rise to bimodal dynamics. Thus, the mechanism of noise-induced bimodality in such systems without CB remains unclear.

\begin{figure}[ht]
    \centering
    \includegraphics[width = 0.62\columnwidth]{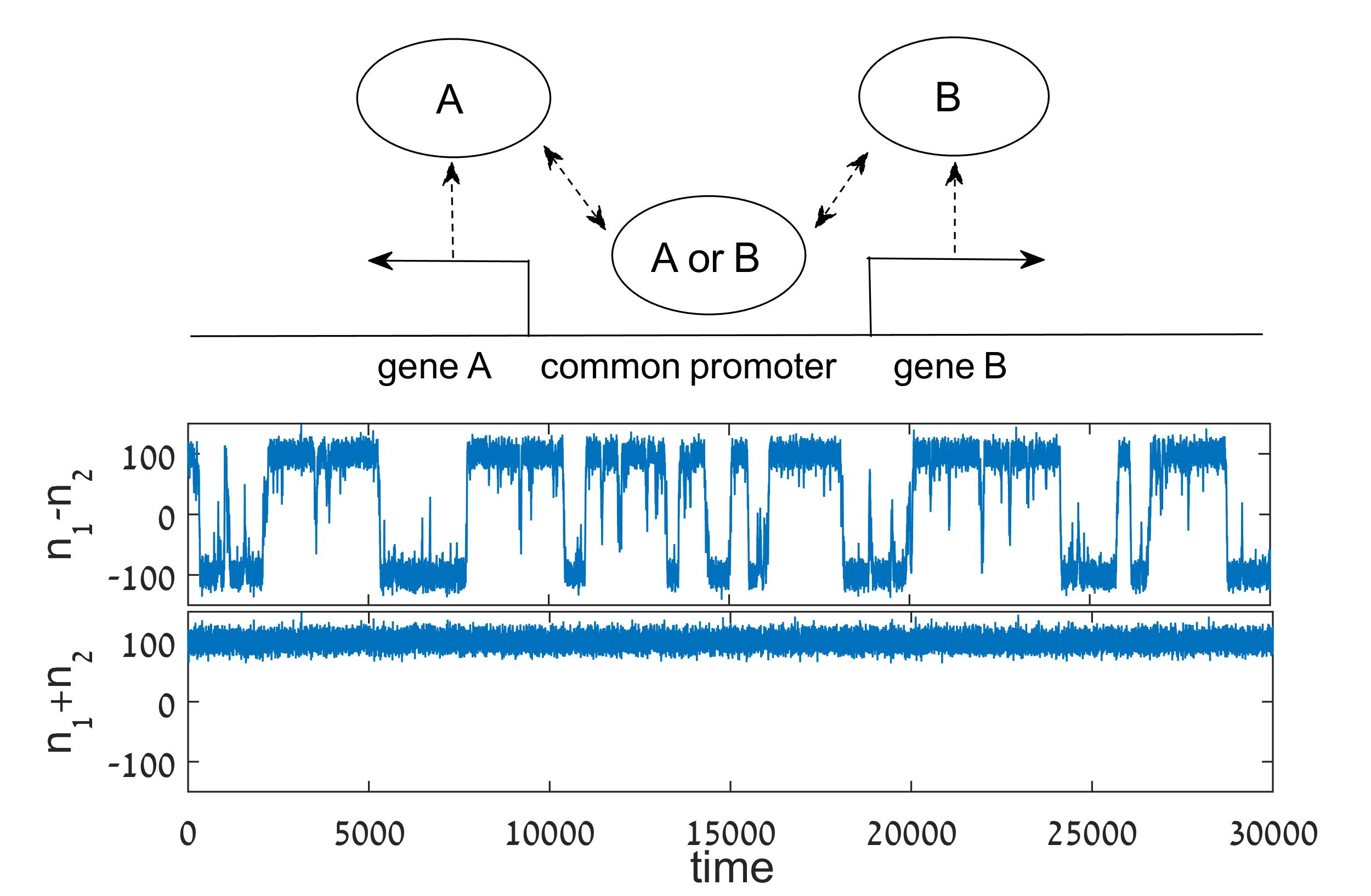}
    \caption{Top: a schematic plot of the ESM~\cite{loinger2007stochastic}. The repressors A and B cannot be bound simultaneously due to overlap between their promoter sites. Middle and bottom: the difference and sum of the copy numbers of $A$ and $B$ obtained from stochastic simulations~\cite{Gillespie1977}, with $\alpha=0.01$, $k=10$ and $g=1$.}
    \label{scheme}
\end{figure}

In this Letter we present an analytical treatment of the ESM, see Fig.~\ref{scheme}, which is found, \textit{e.g.}, as a coarse-grained description of the lysis-lysogeny switch of phage $\lambda$~\cite{ptashne1992phage,golding2011decision}. We begin by analyzing the case of equal degradation rates of the transcription factors. We show that bimodality is driven by multiplicative noise, thus the bimodal states correspond to states for which the noise in the system vanishes. We further find a transition between the bimodal and non-bimodal regimes controlled by the noise strength, and identify the onset of bimodality as function of the repressor strength. Finally, we show that the mean switching time (MST) from a state rich in $A$ to a state rich in $B$ scales polynomially in the system size, unlike typically found in bistable systems. These claims are then generalized to the case of different degradation rates using an adiabatic approximation. Finally, we show that our results hold for other models  displaying noise-induced bimodality such as the general toggle switch~\cite{lipshtat2006genetic, loinger2007stochastic}. Our analysis is also available in the Supplemental Material (SM) and \textit{Mathematica} files.

The genetic toggle switch models mutual inhibition and degradation of transcription factors. In the case of ESM, there is an overlap between the promoters of A and B preventing simultaneous occupation of the two~\cite{allen2005sampling,lipshtat2006genetic}, see Fig.~\ref{scheme}. Thus, at the deterministic level, the dynamics of the free proteins $A$ and $B$, and the bound proteins, $r_A$ and $r_B$, satisfy the following set of equations~\cite{loinger2007stochastic}
\begin{eqnarray} \label{ESMrate}
	\dot{n}_1 & = & g_A(1-r_B)-d_A n_1 - \kappa_0 n_1(1-r_A-r_B)+\kappa_1 r_A\nonumber\\
	\dot{n}_2 & = & g_B(1-r_A)-d_B n_2 - \kappa_0 n_2(1-r_A-r_B)+\kappa_1 r_B\nonumber\\
	\dot{r}_A & = & \kappa_0 n_1(1-r_A-r_B) - \kappa_1 r_A\nonumber\\
	\dot{r}_B & = & \kappa_0 n_2(1-r_A-r_B) - \kappa_1 r_B.
\end{eqnarray}
Here $n_1$ and $n_2$ denote the copy-numbers of proteins $A$ and $B$, respectively. Also, $g_A$ and $g_B$ are the maximal production rates of proteins $A$ and $B$, and $d_A$ and $d_B$, the corresponding degradation rates. In addition, the bound repressors $r_A$ and $r_B$, $0\leq r_A,r_B\leq 1$, are bound $A$ and $B$ proteins that monitor the production of $B$ and $A$, respectively, $\kappa_0$ denotes the binding rate of proteins to the promoter while $\kappa_1$ is the dissociation rate.

For simplicity we will henceforth assume $g_A=g_B=g$. In the limit of $d_A,d_B\ll \kappa_1$, the relaxation of the bound proteins is fast compared to that of the free proteins. As a result, in this limit, one can adiabatically eliminate the fast variables $r_A$ and $r_B$ and arrive at a set of two Michaelis-Menten-like rate equations for $n_1$ and $n_2$~\cite{loinger2007stochastic}:
\begin{equation} \label{nmsys}
\dot{n}_{1} = f_1(n_1,n_2)-\alpha_1 n_1\;,\;\;\;\dot{n}_{2} = f_2(n_1,n_2)-\alpha_2 n_2,
\end{equation}
where $f_i(n_1,n_2)=(1+kn_i)/(1+kn_1+kn_2)$. Here we have defined the dimensionless repression strength $k=\kappa_0/\kappa_1$ as the ratio of the binding and unbinding rates, $\alpha_1=d_A/g$ and $\alpha_2=d_B/g$ are the rescaled degradation rates of $A$ and $B$, and we have rescaled time $t\to gt$. We will further assume that $\alpha_1=\alpha_2\equiv \alpha$, which will be generalized later on.

In this paper we focus on the strong repression limit, $kn_i\gg 1$ ($i=1,2$)~\cite{loinger2007stochastic}, which is found (\textit{e.g.}) in a bacterial genetic switch~\cite{gardner2000construction}. Since at the fixed point of system~\eqref{nmsys} $n_i\sim \alpha^{-1}$, see below,  the strong repression limit becomes $\varepsilon\equiv \alpha/k\ll 1$, and one can naturally define the concentrations of $A$ and $B$ by $x_1=\alpha n_1$, $x_2=\alpha n_2$, respectively. The scaling of the fixed points allows us to introduce the effective system size $\alpha^{-1}$. Yet, while $\alpha^{-1}$ is proportional to the physical system size $N$ originating from system~(\ref{ESMrate}), they are not identical. In the SM we discuss in detail the relationship between our rescaled parameters and the physical system size, and we also comment about the biological relevance of our approximations. Finally, note that at the fixed point, $n_1^*=n_2^*\simeq (1+\varepsilon)/(2\alpha)$, see SM, indicating that, in the deterministic limit, the system converges into an equal state of $A$'s and $B$'s.



To account for demographic stochasticity ignored by Eqs.~\eqref{nmsys}, we can write down the corresponding master equation for the probability $P_{n_1,n_2}$ to find $n_1$ and $n_2$ molecules of type $A$ and $B$, respectively. Defining the step operator $E_n^\pm F(n)=F(n \pm 1)$, we have (see SM):
\begin{eqnarray}\label{auxmaster}
\dot{P}_{n_1,n_2}&=&\left[(E_{n_1}^{-}-1)f_1(n_1,n_2)+(E_{n_2}^{-}-1)f_2(n_1,n_2)\right.\nonumber\\
&+&\left.\alpha_1(E_{n_1}^+-1)n_1+\alpha_2(E_{n_2}^+-1)n_2\right]P_{n_1,n_2}.
\end{eqnarray}
Using the Gillespie algorithm~\cite{Gillespie1977}, stochastic system~(\ref{auxmaster}) is simulated and shown to exhibit bimodality in some range of parameters (middle panel in Fig.~\ref{scheme}), in sharp contrast with the deterministic dynamics~(\ref{nmsys})~\cite{lipshtat2006genetic, loinger2007stochastic}.

To this end, we introduce two auxiliary variables: the total concentration, $w=x_1+x_2$, and the (adimensional) concentration difference $u=(x_1-x_2)/(x_1+x_2)$. Note that $u\approx \pm 1$ when the system is rich in one type of transcription factor, whereas $u\approx 0$ at the deterministic fixed point. For strong repression, $\varepsilon\ll 1$, the joint stationary probability density function (PDF), ${\cal P}_s(u,w)$, decouples and satisfies  ${\cal P}_s(u,w)=P_s(u)R_s(w)$ (see SM). Here
\be
	R_s(w)=(2\pi\alpha)^{-1/2}e^{-\frac{[w-(1+\varepsilon)]^2}{2\alpha}},
\ee
indicating that the sum of $A$'s and $B$'s, represented by $w$, is approximately conserved. To find $P_s(u)$, we consider its Langevin equation (see SM)
\be \label{ueq}
	du/d\tilde{t} = -u + \sqrt{k} \sqrt{1-u^2}\eta(\tilde{t}),
\ee
where $\tilde{t} = 2 g \epsilon \alpha t = 2g \alpha^{2} t /k$, $t$ is the physical time used in~\eqref{ESMrate}, and $\eta(t)$ denotes normalized Gaussian white noise.

\begin{figure}[ht]
    \centering
    \includegraphics[width = 0.6\columnwidth]{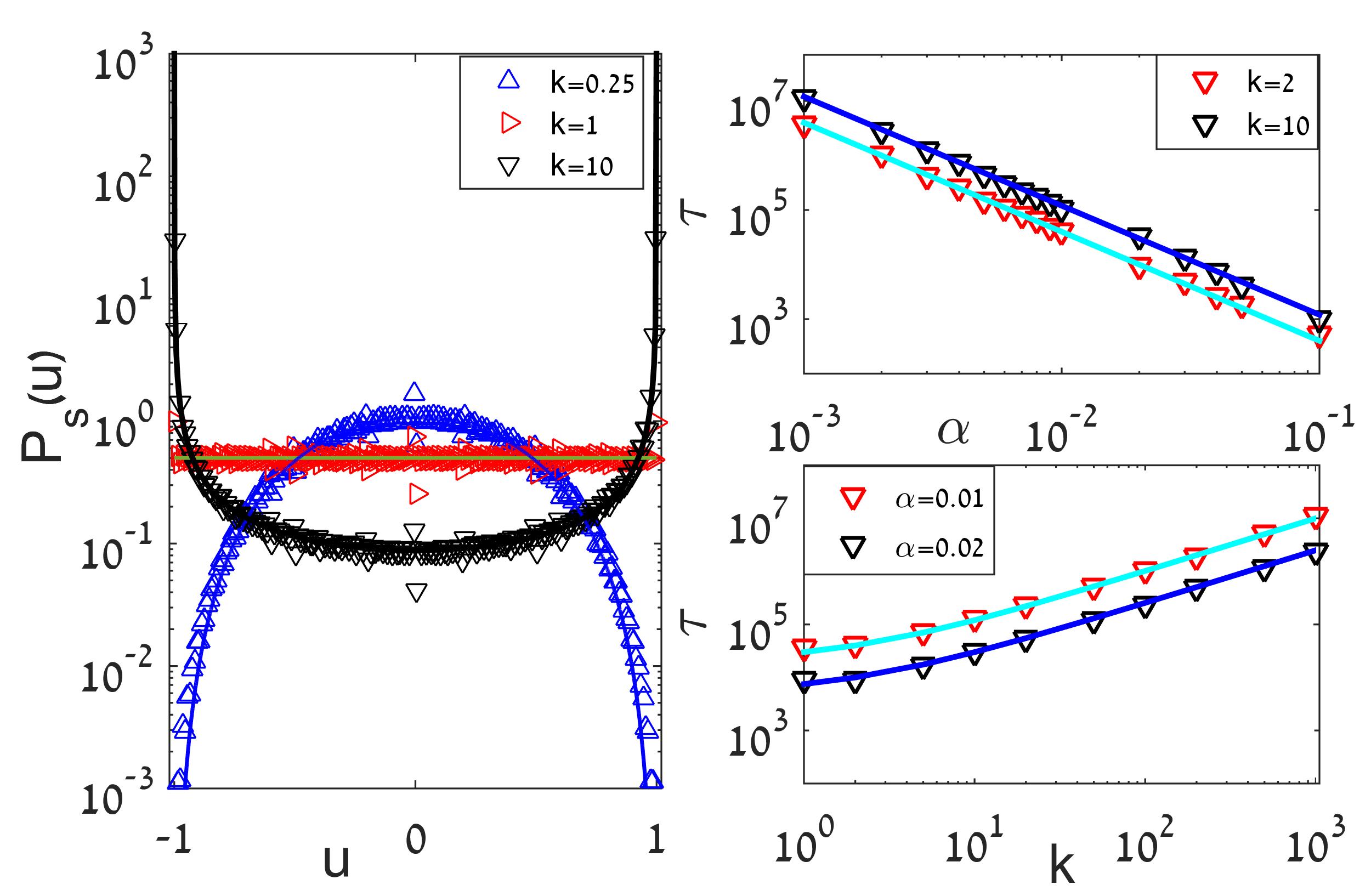}
    \caption{Left panel: The PDF $P_s(u)$ for different values of $k$. For $k>1$, a bimodal PDF appears, for $k=1$ the PDF is flat and, for $k<1$, unimodal with a peak on $u=0$. Solid lines are given by Eq.~\eqref{psu} while markers are obtained by simulations~\cite{Gillespie1977}, with $\alpha = 0.01$. Right panels: The MST as a function of $\alpha$ (upper right panel) and $k$ (lower right panel) for $g=1$. Each marker is obtained by averaging 200 numerical realizations~\cite{Gillespie1977}, whereas solid lines are given by Eq.~\eqref{mst}.}
    \label{pdfs}
\end{figure}

Equation~\eqref{ueq} captures the stochastic dynamics of the system. It has already been treated in previous works~\cite{biancalani2014noise, biancalani2015statistics}, and suggests an explanation for the occurrence of bimodality in the genetic toggle switch. The deterministic drag, $-u$, attracts the system to the stable fixed point, $u^{*}=0$, but since at this state the noise has maximum strength, $\sqrt k$, the value of $u$ is driven away, toward those states at which the noise vanishes, $u=\pm 1$. These are the bimodal states and replace the deterministic fixed points in the CB case.
How does this result depend on the repressor strength $k$? Our previous argument has assumed that the noise strength at fixed point is large enough to oppose the deterministic drag. Yet, taking $k\to 0$, yields $\dot u = -u$, and thus $u(t) \to 0$ as $t\to\infty$. We can thus expect that for small $k$'s, the system fluctuates around $u=0$ without exhibiting bimodality. This transition from unimodality to bimodality is elucidated by the stationary PDF, $P_s(u)$, of Eq.~\eqref{ueq}~\cite{Gardiner2009}. We find
\be \label{psu}
	P_s(u) = \mathcal N \left(1-u^2\right)^{(1-k)/k},
\ee
where $\mathcal N = \Gamma \left(k^{-1}+1/2\right)/[\sqrt{\pi}\,\Gamma \left(k^{-1}\right)]$ is a normalization constant such that $\int_{-1}^{1}P_s(u)du=1$.  Defining the critical repressor strength, $k_C = 1$ (where the PDF concavity is changed), we find  two distinct regimes:  non bimodal, $k<k_{C}$, where the system displays Gaussian fluctuations around the fixed point $u^{*}=0$, and bimodal, $k>k_{C}$, where the system exhibits bimodality and switches between the states $u=\pm 1$. In Fig.~\ref{pdfs}, Eq.~(\ref{psu}) excellently agrees with simulations for different values of $k$. Finally, that PDF~(\ref{psu}) satisfies $|P_s(u+\alpha)-P_s(u)|\ll P_s(u)$, at $u\in(-1,1)$,
validates a-posteriori the Fokker-Planck approximation, see SM, to the master equation~(\ref{auxmaster})~\cite{doering2005extinction,Assaf2007}.

Equation~\eqref{ueq} also allows calculating the MST between the bimodal states~\cite{biancalani2014noise, biancalani2015statistics}.
In the bimodal regime, the MST $\tau$ is the mean time it takes the system to go from a state rich in one transcription factor, say $u=1$, to a state rich in the other, $u=-1$, or vice versa. As shown in~\cite{biancalani2014noise,biancalani2015statistics}, for $k\gg 1$,
the MST of Eq.~\eqref{ueq} reads
\be \label{mst}
	\tau \simeq (k + 2)/(g\alpha^{2}),
\ee
where we have restored the original time units used in~\eqref{ESMrate}. This result (checked against simulations in Fig.~\ref{pdfs}) depends polynomially on the effective system size $\alpha^{-1}$, in contrast with the usually found exponential dependence of the mean escape time in bistable switches, see \textit{e.g.} Refs.~\cite{dykman1994large,Escudero2009,Assaf2010ems,Assaf2011dsg,newby2015bistable}. Hence, the absence of CB allows for much more frequent switching between different phenotypic states, which can be beneficial, \textit{e.g.}, in cases of severe stress~\cite{Balaban2004bpa}.

The previous results can be generalized to the case of different degradation rates, which can be analyzed using an adiabatic elimination of the $w$ variable~\cite{constable2014fast, Kampen2007, Gardiner2009}. A similar treatment can also be used to investigate the case of different repression strengths $k_1\neq k_2$. Yet, as can be checked, for $\varepsilon\ll 1$ the effect of uneven $k$'s on the PDF and MST is much weaker that the effect of uneven $\alpha$'s.

\begin{figure}[htpc!]
    \centering
		\includegraphics[width = 0.6\columnwidth]{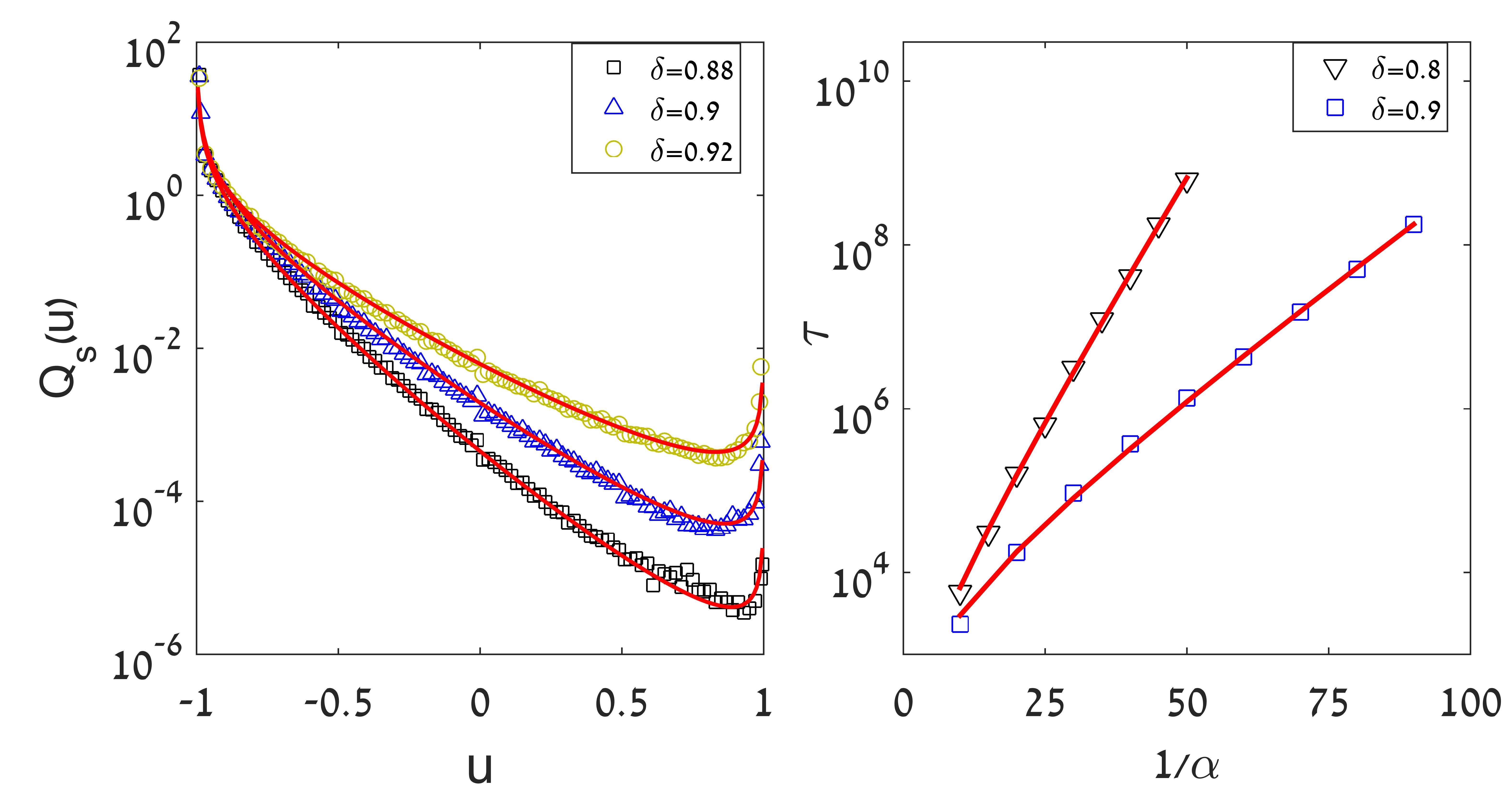}
    \caption{Left panel: $Q_s(u)$ [Eq.~(\ref{pdf})] (solid lines) is compared for different values of $\delta$ against simulations~\cite{Gillespie1977} (symbols). Here $k=5$ and $\alpha=0.01$. Right panel: MST $\tau$ versus $1/\alpha$, for $k=50$ and $g=1$. Each marker is obtained by averaging 200 numerical realizations, while the solid lines are given by Eq.~(\ref{mstdelta}) with $A=50$ for $\delta=0.8$ and $A=100$ for $\delta=0.9$.}
    \label{diffdelta}
\end{figure}

We again consider Eqs.~(\ref{nmsys}) assuming, without loss of generality, $\alpha_2 < \alpha_1$, and denote $\alpha_1\equiv\alpha$ and $\alpha_2\equiv\delta\alpha$, where $\delta \in (0, 1]$. Defining $u=(x_1-x_2)/(x_1+x_2)$ and $w=x_1+x_2$, where $x_1=\alpha n_1$ and $x_2=\alpha n_2$ are the concentrations, the stationary PDF, $Q_s(u)$, of finding concentration $u$, reads (see SM for details)
\begin{eqnarray} \label{pdf}
	&&Q_s(u)=\mathcal Z P_s(u) (1+u+\delta-u\delta)^{-1-\frac{2}{\alpha(1+\delta)}}\nonumber\\
	&& \times \exp \left( \frac{1}{k} \left(\frac{1-\delta}{1+\delta}\right) \left[2u+\ln\left(\frac{1-u}{1+u} \right) \right] \right),
\end{eqnarray}
where $P_s(u)$ is given by Eq.~\eqref{psu}, and $\mathcal Z$ is a normalization factor such that $\int_{-1}^{1}Q_s(u)du=1$. Our theory [Eq.~(\ref{pdf})]
excellently agrees with simulations, see Fig.~\ref{diffdelta}.

The PDF~\eqref{pdf} is a tilted version of PDF~\eqref{psu}; indeed, the former reduces to the latter for $\delta=1$. Since we have chosen $\delta<1$, we
find that the system resides most of the time at the metastable mode of $u=-1$ and occasionally jumps to the transiently metastable mode of $u=1$ (the opposite
would occur for $\delta>1$). Similarly as for the case of $\delta=1$, by decreasing $k$ there exists a transition from a state rich in one type of
transcription factor to a state where both types coexist, although not equally. Again, this is determined by a critical repressor strength $k_C$, satisfying $k_C = 2/(1+\delta)$, see SM.
For $k>k_C$, both $u=1$ and $u=-1$ are noise-induced metastable states, although the system is biased toward $u=-1$ as the degradation rate of the
corresponding protein (of type $B$) is smaller. In contrast, as $k$ is decreased below $k_C$, the PDF flips, and peaks at $u^*=-1+{\cal O}(\varepsilon)$, see SM.

Since the MST $\tau$ from $u=-1$ to $u=1$ turns out to depend exponentially on the effective system size $\alpha^{-1}$ (see below), given Eq.~(\ref{pdf}), $\tau$ satisfies in the leading order $\tau\sim Q_s(-1)/\min[Q_s(u)]$~\cite{dykman1994large,Assaf2006stm}. Here, the minimum of $Q_s(u)$ is obtained in the close vicinity of $u=1$, satisfying $u_m\simeq 1-2\varepsilon(k/k_C-1)/(1-\delta)\simeq 1$. As $Q_s(u)$ diverges at $u=-1$, we thus compute the limit $\lim_{a\to 0} Q_s(-1+a)/Q_s(u_m)$ and find, in the leading order of $\varepsilon\ll 1$
\begin{equation}\label{mstdelta}
\tau\simeq \frac{\mathcal A}{g\alpha}\exp\left[\frac{2}{\alpha(1+\delta)}\ln\frac{1}{\delta}\right].
\end{equation}
Here $\mathcal{A} = \mathcal{A}(k,\delta)$ is an unknown prefactor, and we have restored the physical time units.
Equation~(\ref{mstdelta})
agrees well with simulations, see Fig.~\ref{diffdelta}, and in contrast to Eq.~(\ref{mst}), depends exponentially on the effective system size.

\begin{figure}[htpc!]
    \centering
    		\includegraphics[width = 0.65\columnwidth]{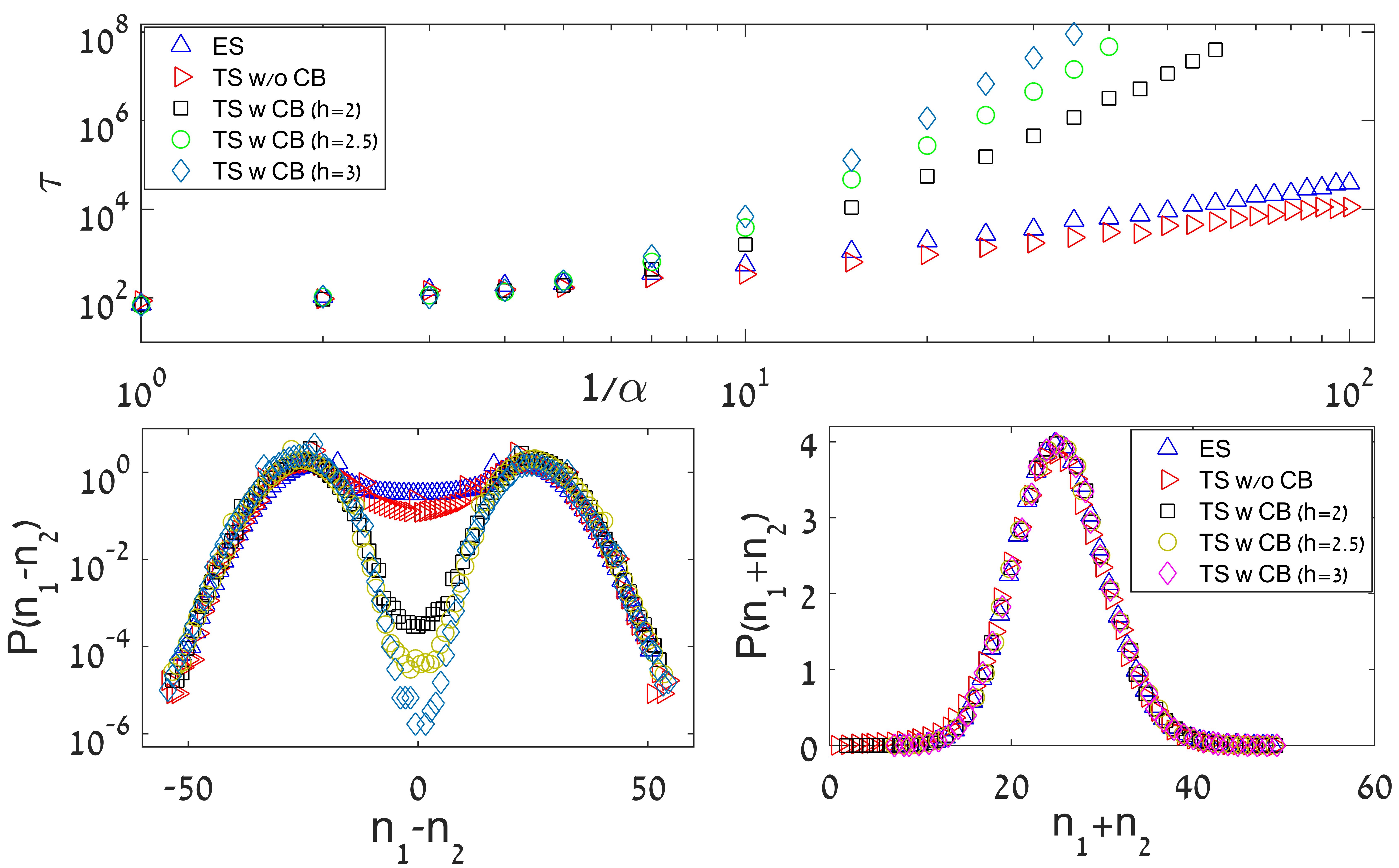}
    \caption{(Top) MSTs for five different models: ESM, general toggle switch (TS) without (w/o) CB, and TS with (w) CB with $h=2,2.5,3$, for $k=1.5$. Each point is obtained by averaging 200 realizations. (Bottom) PDFs of the difference and sum of the copy numbers $n_1$ and $n_2$, for $k=5$ and $\alpha=0.04$.
    While $P(n_1+n_2)$ almost coincides for all models, the ``potential barrier"
    for switching given by $\max[P(n_1-n_2)]-\min[P(n_1-n_2)]$, is much shallower for models without CB. }
    \label{difftau}
\end{figure}

Finally, we can use the analysis above for other models that exhibit noise-induced bimodality such as the general toggle switch,
described by Eqs.~(\ref{nmsys}) with
\begin{equation}\label{fi}
f_i(n_1,n_2)=[1+(kn_j)^h]^{-1}\;,\;\;\;i\neq j=1,2
\end{equation}
where the Hill coefficient is $h=1$~\cite{loinger2007stochastic}. In principle, the analysis can be done in the same manner as for the ESM. Yet, the task is slightly more difficult since the Langevin equation for $w=x_1+x_2$ does not yield a Gaussian PDF for $R_s(w)$, which makes the equation for $u$ less tractable. Nonetheless, we have numerically found the onset of bimodality to be at $k>k_C=1$ and that the MST behaves similarly to the ESM, see Fig.~\ref{difftau}. In sharp contrast, the genetic toggle switch model \textit{with} CB, for which $f_i(n_1,n_2)$ are given by Eq.~(\ref{fi}) with Hill coefficient $h\geq 2$, displays (at least) two stable fixed points. In this case there is no threshold for bimodality when $\varepsilon\ll 1$, and one expects an exponential dependence of the MST on the system's size~\cite{newby2012isolating}. In Fig.~\ref{difftau} we compare the MSTs and PDFs of several models with and without CB. Our simulations indicate that the MST in the case of CB with $h\geq 2$ yield a stretched-exponential dependence of the MST on the system's size. This is a nontrivial result and requires a further study. While this is beyond the scope of this paper, we believe the formalism we have developed can be used to study toggle switch models with CB as well, as long as we are in the strong repression limit.

We have presented an analytical treatment of the ESM demonstrating a bimodal behavior in the absence of two stable fixed points at the deterministic level.
Bimodality is induced by multiplicative noise: the noise strength vanishes at the bimodal states whereas it is maximal at the single stable fixed point. This phenomenon, which has attracted much interest in various
fields~\cite{Togashi2001,doering1986stability, artyomov2007purely, Ohkubo2007, remondini2013analysis, biancalani2014noise}, is linked here to previous
numerical~\cite{lipshtat2006genetic, loinger2007stochastic} and experimental~\cite{gardner2000construction} findings on the genetic toggle switch.

We have shown that bimodal behavior ceases to occur if the noise strength in the system, controlled by the repression strength $k$, is reduced below a critical threshold. This transition, absent in bistable systems, is similar to that found in other noise-induced bimodal systems~\cite{Togashi2001, russell2011noise, biancalani2014noise}. Moreover, we have shown here that the MST between bimodal states exhibits a polynomial, rather than exponential, scaling on the system size. In genetic toggle switches, the noise is controlled by the repression
strength $k$, suggesting that bimodality can be achieved or lost by biological fine tuning of reaction rates.


We would like to thank Ofer Biham for valuable discussions. This work was supported by Grant No.~300/14 of the Israel Science Foundation.  T.B. acknowledges partial support from the National Aeronautics and Space Administration through the NASA Astrobiology Institute under Cooperative Agreement No. NNA13AA91A issued through the Science Mission Directorate.


\newpage 
\setcounter{equation}{0}
\renewcommand{\theequation}{S\arabic{equation}}
\begin{center}
{{\itshape\Large Supplemental Material for:}\\~\\\Large Genetic Toggle Switch in the Absence of Cooperative Binding: Exact Results}\\~\\
{\large Tommaso Biancalani and Michael Assaf}\\
\end{center}

\subsection{The exclusive switch model in the case of equal degradation rates}
\subsubsection{Model definition and deterministic dynamics}
Our starting point here are rate equations (2) in the main text, obtained in the limit of fast binding/unbinding compared to other processes in the circuit~\cite{loinger2007stochastic}, see main text. These equations describe the dynamics of the mean number of $A$ and $B$ proteins, denoted by $n_1$ and $n_2$, respectively, and read
\begin{equation} \label{sm_odes}
	\dot{n}_1 =  \frac{1+kn_1}{1+kn_1+kn_2} - \alpha_1 n_1,\quad \dot{n}_2 = \frac{1+kn_2}{1+kn_1+kn_2} - \alpha_2 n_2.
\end{equation}
Here, we have denoted the repression strength by $k=\kappa_0 / \kappa_1$, which is the ratio between the binding rate $\kappa_0$ and unbinding rate $\kappa_1$ to/from the promoter. In addition,  we have taken $g_A=g_B\equiv g$, rescaled time $t\to gt$, and denoted the rescaled degradation rates $\alpha_1=d_A/g$ and $\alpha_2=d_B/g$, see main text for the definition of parameters.

System~\eqref{sm_odes} admits a unique positive stable fixed point. Assuming $\alpha_1=\alpha_2\equiv \alpha$ (we will relax this assumption later on), we find
\begin{equation} \label{nfp}
	n^*_1 = n^*_2 = \frac{k -\alpha +\sqrt{\alpha ^2+k^2+6 \alpha  k}}{4 k \alpha}.
\end{equation}
From now on we will consider the strong repression limit, $\varepsilon\equiv\alpha/k \ll 1$, meaning that degradation is assumed to occur much slower than inhibition. In this limit, the fixed point~\eqref{nfp} simplifies to $n^*_1 = n^*_2 \simeq 1/(2\alpha)$. As a result, in the strong repression limit we can naturally define an \textit{effective} system size as $\alpha^{-1}$. By additionally defining the protein concentrations $x_i = \alpha n_i$, $i=1,2$, the rate equations~(\ref{sm_odes}) become
\begin{equation}\label{REs}
	\dot{x}_1=\frac{x_1+\varepsilon }{x_1+x_2+\varepsilon }- x_1, \quad \dot{x}_2=\frac{x_2+\varepsilon }{x_1+x_2+\varepsilon }- x_2,
\end{equation}
where we have further rescaled time by $t\to \alpha t$. Equations~(\ref{REs}) are governed by a single parameter, $\varepsilon$, although $\alpha$ still appears in the definition of the concentrations, a fact that we shall use to carry out the expansion of the master equation.

\subsubsection{Relationship between model parameters and physical system size and biological relevance of parameter values}
As discussed in the main text, system~\eqref{sm_odes} is obtained by applying an adiabatic approximation of the following system of ODEs:
\begin{eqnarray} \label{sm_ESMrate}
	\dot{n}_1 & =& g_A(1-r_B)-d_A n_1 - \kappa_0 n_1(1-r_A-r_B)+\kappa_1 r_A\nonumber\\
	\dot{n}_2 & =& g_B(1-r_A)-d_B n_2 - \kappa_0 n_2(1-r_A-r_B)+\kappa_1 r_B\nonumber\\
	\dot{r}_A & =& \kappa_0 n_1(1-r_A-r_B) - \kappa_1 r_A\nonumber\\
	\dot{r}_B & =& \kappa_0 n_2(1-r_A-r_B) - \kappa_1 r_B,
\end{eqnarray}
where $g_A = g_B = g$ are the maximal production rates of proteins $A$ and $B$, $d_A=d_B=d$ are the corresponding degradation rates, and $r_A$ and $r_B$ the copy numbers of bound repressors. The constant $\kappa_0$ denotes the binding rate of proteins to the promoter while $\kappa_1$ is the dissociation rate. Recall also that these parameters are related to those of system~\eqref{sm_odes} via $k=\kappa_0/\kappa_1$ and $\alpha=d/g$.

Let us denote the physical system size by $N$. Changing the physical system size signifies using a larger reservoir with a larger number of molecules, but without changing the rules between the single interacting molecules. Since some of the terms in system~\eqref{sm_ESMrate} depend on the number of molecules in play (and some others do not), it is the purpose of this section to identify how the parameters in system~\eqref{sm_ESMrate} scale with respect to $N$. Clearly, we expect according to the definition of the system size that $n_i \sim N$ and also, that the degradation rates $d_A$ and $d_B$ do not depend on $N$.

The adiabatic reduction assumes that the variables $r_A$ and $r_B$ evolve faster than $n_1$ and $n_2$. This indicates that the terms on the right hand side of the third and fourth equations of system~\eqref{sm_ESMrate} approximately balance each other, and we find \textit{e.g.}
\begin{equation}
		\kappa_0 n_1(1-r_A-r_B) \sim \kappa_1 r_A.
\end{equation}

As a result, since $r_A, r_B\sim \mathcal O(1)$, we find that $k = \kappa_0 / \kappa_1 \sim N^{-1}$. Moreover, close to the fixed point, one can follow a similar argument for the first and second equations of system~(\ref{sm_ESMrate}), and find that $g \sim N^{-1}$, while $\alpha \sim N^{-1}$. Therefore, both the repression strength constant $k$ and the effective degradation rate $\alpha$ are inversely proportional to the physical system size $N$. Yet, in our rescaled model, because of the scaling of the fixed point~\eqref{nfp} on $\alpha$, it is legitimate to define the concentrations variables by $x_i = \alpha n_i$ ($i=1,2$), rather than $x_i = N^{-1} n_i$. Note, that for a given $N$, one can increase/decrease the effective system size $\alpha^{-1}$ by changing the physical degradation rate $d$, while keeping $k$ constant.

We now discuss the biological relevance of the approximations used throughout the paper. We make use of three assumptions: (i) $k$ is assumed to be on the order of one; (ii) the strong repression limit implies that $\epsilon = k/\alpha \ll 1$; (iii) $\alpha$ is assumed to be small so that we can expand the master equation in $\alpha\ll 1$.

Assumption (i) can always be obtained by rescaling time and the copy number variables. If assumption (i) is satisfied, then assumption (iii) follows from (ii). Therefore, we only need to justify assumption (ii) which is the strong repression limit $\varepsilon\ll 1$. As stated, this limit assumes that protein degradation occurs much slower than inhibition. Indeed, this assumption has often been invoked in previous theoretical works~\cite{lipshtat2006genetic, loinger2007stochastic}. Moreover, in Ref.~\cite{gardner2000construction}, the genetic switch model is compared to an experimental switch engineered using \textit{E. Coli}. It is found that in order to have bimodality, the inhibition rate should be $15$ times faster than the degradation rate in one species, and $150$ times faster in the other (see caption of Fig.~5 in Ref.~\cite{gardner2000construction}). This number provides realistic values for $\varepsilon^{-1}$. Furthermore, a similar value of $\varepsilon=0.04$ appears in Ref.~\cite{allen2005sampling}.

\subsubsection{Stochastic dynamics and Fokker-Planck/Langevin approximation}
The stochastic exclusive switch model can be obtained by viewing the terms of system~\eqref{sm_odes} as microscopic rates of production and degradation of transcription factors. Let us denote by $T_1^+$ and $T_2^+$ the production rates, and by $T_1^-$ and $T_2^-$ the degradation rates of proteins of type $A$ and $B$, respectively. Given a state with $n_1$ and $n_2$ particles of type $A$ and $B$ respectively, these transition rates satisfy
\begin{eqnarray} \label{sm_trates}
	T_1^+(n_1,n_2) =  \frac{1+ k n_1}{1+k n_1+k n_2} , & \quad T_1^-(n_1,n_2) =  \alpha n_1,\nonumber\\
	T_2^+(n_1,n_2) =  \frac{1+ k n_2}{1+k n_1+k n_2} , & \quad T_2^-(n_1,n_2) =  \alpha n_2.
\end{eqnarray}
Using these, we can write down the master equation for the evolution of the probability density function (PDF) $P_{n_1, n_2}(t)$ that the system is in state $(n_1, n_2)$ at time $t$:
\begin{eqnarray} \label{sm_me}
    \dot{P}_{n_1,n_2}=T_1^+(n_1-1,n_2)P_{n_1-1,n_2}-T_1^+(n_1,n_2)P_{n_1,n_2}+T_2^+(n_1,n_2-1)P_{n_1,n_2-1}-T_2^+(n_1,n_2)P_{n_1,n_2}\nonumber \\
    +T_1^-(n_1+1,n_2)P_{n_1+1,n_2}-T_1^-(n_1,n_2)P_{n_1,n_2}+T_2^-(n_1,n_2+1)P_{n_1,n_2+1}-T_2^-(n_1,n_2)P_{n_1,n_2}.
\end{eqnarray}
Equation~\eqref{sm_me} with transition rates~\eqref{sm_trates} defines the stochastic model for the case of equal degradation rates, and can be simulated using the Gillespie algorithm~\cite{Gillespie1977}.

Let us now approximate master equation~(\ref{sm_me}) into a Fokker-Planck equation. We do so by Taylor expanding
\begin{equation}
T_1^\pm(n_1\mp 1,n_2)P_{n_1\mp 1,n_2}\simeq T_1^\pm(n_1,n_2)P_{n_1,n_2}\mp\partial_{n_1}[T_1^\pm(n_1,n_2)P_{n_1,n_2}]+\frac{1}{2}\partial_{n_1}^2[T_1^\pm(n_1,n_2)P_{n_1,n_2}],
\end{equation}
and same for $T_2^\pm$. We now move to the concentration variables $x_i=\alpha n_i$, $i=1,2$, and use the fact that $\partial_{n_i}=\alpha\partial_{x_i}$, which turns the Taylor expansion into a system size expansion. Employing this system size expansion on master equation~(\ref{sm_me}), valid when the system size is large $1/\alpha\gg 1$, we arrive at the following Fokker-Planck equation for $P(x_1,x_2,t)$ -- the probability to find concentrations $x_1$ and $x_2$ of $A$ and $B$ proteins, respectively, at time $t$:
\begin{equation} \label{sm_fpe}
	\partial_{\alpha t}  P(x_1,x_2,t) =  \left[ - \partial_{x_1} \mathcal A_1 - \partial_{x_2} \mathcal A_2 + \frac{\alpha}{2} \sum_{i,j=1}^2 \partial_{x_i} \partial_{x_j} \mathcal B_{ij} \right] P(x_1,x_2,t).
\end{equation}
Here we have neglected  $\mathcal{O}(\alpha^3)$ terms or higher, time is measured in $\alpha t$ units, and
\begin{equation}
	\mathcal A_i = \frac{x_i+\varepsilon }{x_1+x_2+\varepsilon }-x_i,\quad \text{and}\quad \mathcal B_{ij} = \left(\frac{x_i+\varepsilon }{x_1+x_2+\varepsilon }+x_i \right) \delta_{ij}.
\end{equation}
Equation~\eqref{sm_fpe} is equivalent to the system of Langevin equations defined in the It\={o} sense~\cite{Gardiner2009}:
\begin{equation} \label{sm_xieq}
	\begin{split}
	 \frac{dx_i }{d(\alpha t)} &= \mathcal A_i + \sqrt \alpha \sqrt{\mathcal B_{ii}} \eta_i(\alpha t), \quad i=1,2,
	\end{split}
\end{equation}
where $\eta_i(t)$, $i=1,2$, are independent normalized white Gaussian noises.

As a final remark, note that system~\eqref{sm_xieq} is a better approximation to the master equation than the linear noise approximation obtained, e.g, with the van Kampen expansion~\cite{Kampen2007}. In fact, system~\eqref{sm_xieq} captures the multiplicative nature of the noise which, as we shall show, is the main driver of bimodality. The expansion technique that we have used is described in more detail  in~\cite{McKane2013}.

\subsubsection{Simplifying the Langevin equations}
We now introduce the new variables
\begin{equation} \label{sm_wu}
	w = x_1+x_2,\quad u=\frac{x_1-x_2}{x_1+x_2},
\end{equation}
and change variables in the Fokker-Planck equation~\eqref{sm_fpe} or, equivalently, in system~\eqref{sm_xieq} using the It\={o} formula~\cite{Gardiner2009} (the latter is simpler). Doing so, and neglecting $\mathcal O(\sqrt \varepsilon)$ terms in the noise coefficients, we arrive at
\begin{equation} \label{sm_wsm_ueq}
	 \frac{dw}{d(\alpha t)} = \frac{-w^2-w \varepsilon +w+2 \varepsilon }{w+\varepsilon } + \sqrt \alpha\sqrt{1+w}\, \eta_{1}(\alpha t), \quad \frac{du}{d(\alpha t)} = - 2 u \varepsilon \frac{w-\alpha}{w^2 (w+\varepsilon )} + \sqrt \alpha\sqrt{(1-u^2)\frac{1+w}{w^2}}\, \eta_{2}(\alpha t). \\
\end{equation}{}

Following the main text, we indicate by $\mathcal P_s(u,w)$ the stationary PDF of Eqs.~\eqref{sm_wsm_ueq}. Equations~\eqref{sm_wsm_ueq} need to be decoupled in order to be analytically tractable.  Simulations (lower panel of Fig. 1 in the main text) indicate that the total concentration $w$ fluctuates around the only stable fixed point $\left(1-\varepsilon+\sqrt{1+6 \varepsilon +\varepsilon ^2}\right)/2\,\simeq 1+\varepsilon$, a fact we wish to exploit for decoupling the two variables. We can see this effect by Taylor expanding in $\varepsilon\ll 1$ the deterministic part of the first of Eqs.~\eqref{sm_wsm_ueq},
\begin{equation}
	\frac{-w^2-w \varepsilon +w+2 \varepsilon }{w+\varepsilon}  \approx w(w-w^*),\quad \text{where } w^*=1+\varepsilon.
\end{equation}
Hence, it is legitimate to linearize the first of Eqs.~\eqref{sm_wsm_ueq} around $w^*$ which, in the leading order of $\varepsilon\ll 1$, becomes
\begin{equation}
	\frac{dw}{d(\alpha t)} =-(w-w^*)+\sqrt{2\alpha}\,\eta_1(\alpha t).
\end{equation}

Therefore, the stationary PDF for finding concentration $w$ reads
\begin{equation}
	R_s(w)=\frac{1}{\sqrt{2\pi\alpha}} \exp{ \left(-\frac{[w-(1+\varepsilon)]^2}{2\alpha}\right)}.
\end{equation}
This indicates that for $\varepsilon\ll 1$, the total number of $A$'s and $B$'s, represented by $w$, is approximately conserved.

We now approximate the second of Eqs.~\eqref{sm_wsm_ueq} by setting $w$ equal to $w^*$. We further simplify this equation by noting that
\begin{equation} \label{sm_sim}
	\frac{w^*-\alpha}{w^{*2}(w^{*}+\varepsilon)}\approx 1,\quad \text{and}\quad \frac{w^*+1}{w^{*2}} \approx 2,
\end{equation}
and by rescaling time by $2 \varepsilon$. As a result, the Langevin equation for $u$ reads
\begin{equation} \label{sm_ueq}
	\frac{du}{d\tilde{t}} = -u + \sqrt{k} \sqrt{1-u^2}\,\eta_2(\tilde{t}),
\end{equation}
where $\tilde{t} = 2 \varepsilon \alpha t = (2 \alpha^{2}/k) t$, and $t$ is the time used in Eq.~\eqref{sm_me}. Note that in order to reach Eq.~(\ref{sm_me}) we have already rescaled time $t\to gt$, so the physical time units contain an additional $1/g$ factor, see main text. We denote the stationary PDF of Eq.~(\ref{sm_ueq}) by $P_s(u)$. We have thus established that $\mathcal P_s(u,w) = P_s(u)R_s(w)$. Eq.~(\ref{sm_ueq}) and its stationary solution are further discussed in the main text, where we find that the mean switching time scales polynomially with $\alpha^{-1}$. Yet, verifying this result experimentally is expected to be highly nontrivial, as trapping of single cells over many cell cycles is required (\textit{e.g.} by using a microfluidic device) to carry our such measurements.

\subsection{The exclusive switch model in the case of different degradation rates}
\subsubsection{Model definition and deterministic dynamics}
The exclusive switch model~\eqref{sm_odes} can be generalized to the case of different degradation rates:
\begin{equation} \label{sm_odes2}
	\dot{n}_1=\frac{1+ k n_1}{1+k n_1+k n_2}-\alpha_1 n_1,\quad \dot{n}_2=\frac{1+ k n_2}{1+k n_1+k n_2}-\alpha_2 n_2.
\end{equation}
We assume, without loss of generality, $\alpha_2 < \alpha_1$, and denote $\alpha_1\equiv\alpha$ and $\alpha_2\equiv\delta\alpha$, where $\delta \in (0, 1]$. The concentrations of $A$ and $B$ are given respectively by $x_1=\alpha n_1$ and $x_2=\alpha n_2$. Rescaling time $t\to \alpha t$ and using $\varepsilon = \alpha/k$, we arrive at the following rescaled rate equations
\begin{equation} \label{sm_xysys}
	\dot{x_1} = \frac{x_1 + \varepsilon}{x_1+x_2+\varepsilon}-x_1,\quad \dot{x_2} = \frac{x_2 + \varepsilon}{x_1+x_2+\varepsilon}-\delta x_2.
\end{equation}

The corresponding stochastic model is described by a master equation given by Eq.~(\ref{sm_me}) with rates~(\ref{sm_trates}) where here $T_2^-(n_1,n_2)=\delta \alpha n_2$.
Note that in this case we keep our original system size definition as $\alpha^{-1}$.

\subsubsection{Derivation of the Langevin equations and analysis}
Expanding master equation~\eqref{sm_me} in a similar way to the previous section, we arrive at the following system of Langevin equations:
\begin{eqnarray}\label{sm_xieq2}
	\frac{d x_1}{d (\alpha t)} &=& \frac{x_1+\varepsilon }{x_1+x_2+\varepsilon }-x_1 + \sqrt{\alpha } \sqrt{\frac{x_1+\varepsilon }{x_1+x_2+\varepsilon }+x_1}\; \eta_1(\alpha t),\nonumber\\
	\frac{d x_2}{d (\alpha t)} &=& \frac{x_2+\varepsilon }{x_1+x_2+\varepsilon } -\delta  x_2 + \sqrt{\alpha } \sqrt{\frac{x_2+\varepsilon }{x_1+x_2+\varepsilon }+\delta  x_2}\; \eta_2(\alpha t).
\end{eqnarray}
These equations reduce to system~\eqref{sm_xieq} for $\delta=1$.

We now change variables in system~\eqref{sm_xieq2} using the It\={o} formula~\cite{Gardiner2009}, from $(x_1, x_2)$ to $(w, u)$ defined in Eqs.~\eqref{sm_wu}. We obtain
\begin{equation} \label{sm_wu2}
	\frac{dw}{d(\alpha t)} = \mathcal A_w + \mathcal G_{11} \eta_1(\alpha t) + \mathcal G_{12} \eta_1(\alpha t),\quad
	\frac{du}{d(\alpha t)} = \mathcal A_u + \mathcal G_{21} \eta_1(\alpha t) + \mathcal G_{22} \eta_2(\alpha t),
\end{equation}
where the deterministic part for $w$ reads
\begin{equation}
	\mathcal A_w = \frac{1}{2}\left(\delta  u w-u w-\delta  w-\frac{2 w}{w+\varepsilon }-w+4\right),
\end{equation}
and the deterministic part for $u$, written as a term independent of $\delta$ plus a correction, is
\begin{equation} \label{sm_au}
	\mathcal A_u = - 2 u \varepsilon \frac{w-\alpha}{w^2 (w+\varepsilon )} + (\delta - 1) \frac{\left(1-u^2\right) w^2 (\alpha +w)- u^2 w \varepsilon  (\alpha +w)+w \varepsilon  (\alpha +w)}{2 w^2 (w+\varepsilon )}.
\end{equation}
In addition, the expression for the noise matrix $\mathcal G$, neglecting $\mathcal O(\sqrt \varepsilon)$ terms, reads
\begin{equation}
	\mathcal G = \frac{\sqrt \alpha}{w\sqrt 2}\left(\begin{array}{cc}
 	\sqrt{w^2 (u+1) (w+1)} 		& \sqrt{w^2 (1-u) (w \delta +1)} \\
 	(1-u) \sqrt{(1+u) (w+1)} 	& -(1+u) \sqrt{(1-u) (w \delta +1)} \\
	\end{array}\right).
\end{equation}
As a final remark, note that instead of matrix $\mathcal G$ one can just reabsorb the noise coefficients by defining new noise variables, $\xi_1$ and $\xi_2$, so that Eqs.~\eqref{sm_wu2} become
\begin{equation} \label{sm_wu3}
	\frac{dw}{d(\alpha t)} = \mathcal A_w + \xi_1(\alpha t),\quad
	\frac{du}{d(\alpha t)} = \mathcal A_u + \xi_2(\alpha t).
\end{equation}
Here, the correlator of the two noise variables reads $\langle \xi_i(t)\xi_j(t')\rangle = \mathcal B_{ij}(u,w) \delta(t-t')$, with $\mathcal B = \mathcal G \mathcal G^T$. This is indeed the result that one arrives at by applying the change of variable~\eqref{sm_wu} to the Fokker-Planck equation rather that using the It\={o} formula on the corresponding Langevin system. Yet, we have pursued the latter approach since it is simpler.

In order to decouple Eqs.~\eqref{sm_wu2}, we compute the single $u$-dependent positive fixed point of the $w$ equation, $w^*(u)$. Taylor expanding this fixed point in $\varepsilon\ll 1$, we arrive at the expression:
\begin{equation} \label{sm_adiab}
	w^*(u) \approx \frac{2}{\delta -\delta  u+u+1}.
\end{equation}
We now employ an adiabatic elimination~\cite{Assaf2008noise} of $w$ by substituting $w= w^*(u)$ into the $u$ equation. Moreover, we sum the two noise terms using the summation rule for Gaussian variables~\cite{Gardiner2009}
\begin{equation}
	\mathcal G_{21} \eta_1(\alpha t) + \mathcal G_{22} \eta_2(\alpha t) = \sqrt{\mathcal G_{21}^2 + \mathcal G_{22}^2}\, \eta(\alpha t) = \sqrt{\mathcal B_{uu}} \eta(\alpha t),\quad \text{where }\; \mathcal B_{uu} = \alpha\left(1 - u^2\right) \left\{\frac{w^*(u) [\delta +(\delta -1) u+1]+2}{2 w^*(u)^2}\right\},
\end{equation}
and $\eta(t)$ is normalized white Gaussian noise. Substituting Eq.~\eqref{sm_adiab} into the expression for $\mathcal B_{uu}$, we arrive at the simple formula
\begin{equation}
	\mathcal B_{uu} = \frac{\alpha}{2}(1-u^2) (1+\delta) [\delta +(1-\delta)u +1].
\end{equation}
We now simplify the deterministic part of the $u$-equation, Eq.~\eqref{sm_au}, by neglecting small terms. The first term of Eq.~\eqref{sm_au} is simplified according to  Eq.~\eqref{sm_sim}, as done in the previous section. The second $\delta$-dependent term is approximated by taking the leading order result with respect to $\alpha\ll 1$ and $\varepsilon\ll 1$. As a result, we arrive at the final Langevin equation for $u$:
\begin{equation}\label{Langeu}
	\frac{du}{d(\alpha t)} = \frac{1}{2} (\delta -1) \left(1-u^2\right)-\frac{1}{2} u \varepsilon(\delta-\delta u +u+1)^2+  \sqrt{\frac{\alpha}{2}}\sqrt{(1-u^2)(1+\delta) [\delta +(1-\delta)u +1]}\;\eta(\alpha t).
\end{equation}
This equation describes the model for the case of different degradation rates and its stationary solution is discussed in the main text. The stationary PDF, $Q_s(u)$, for finding concentration $u$ can be found from Eq.~(\ref{Langeu}) and reads
\begin{equation} \label{pdfSM}
	Q_s(u)=\mathcal Z \left(1-u^2\right)^{\frac{1-k}{k}} (1+u+\delta-u\delta)^{-1-\frac{2}{\alpha(1+\delta)}}\exp \left( \frac{1}{k} \left(\frac{1-\delta}{1+\delta}\right) \left[2u+\ln\left(\frac{1-u}{1+u} \right) \right] \right),
\end{equation}
where $\mathcal Z$ is a normalization factor such that $\int_{-1}^{1}Q_s(u)du=1$.

\subsubsection{Investigation of the tilted PDF in the case of different degradation rates}
PDF~\eqref{pdfSM} is a tilted version of the PDF in the equal $\alpha$ case, see Eq.~(6) in the main text. Indeed, the former reduces to the latter for $\delta=1$. To remind the reader, without loss of generality we assume $\delta<1$.
When $k>k_C$ (where $k_C$ is the critical repression strength below which bimodality is lost, see below), both $u=1$ and $u=-1$ are noise-induced metastable states. For $\delta<1$, we find that the system resides most of the time in the metastable mode of $u=-1$ and occasionally jumps to the transiently metastable mode of $u=1$ (the opposite would occur for $\delta>1$). This is because the degradation rate of protein $B$ is smaller in this case. In contrast, as $k$ is decreased below $k_C$, first the mode at $u=1$ and then the mode at $u=-1$ are lost, and eventually the PDF flips, and becomes a unimodal PDF with a peak at $u^*$ which is the mode of Eq.~(\ref{pdfSM}). Differentiating the logarithm of PDF~(\ref{pdfSM}) once and equating to zero, we find the PDF mode at $k<k_C$ to be at
\begin{equation}\label{mode}
	u^*\simeq -1+\frac{\delta(\delta+1)\alpha}{k(1-\delta)}\left(\frac{2\delta}{\delta+1}-k\right).
\end{equation}

In order to find $k_C$ in the case of different degradation rates we evaluate $k$ at which $Q_s(u)$ ceases to be bimodal. As $k$ is decreased, the mode at $u=1$ disappears before the mode at $u=-1$ does. As a result, $k_C$ can be found by checking when the concavity at $u=1$ of $Q_s(u)$ changes sign. Alternatively, since for $k>k_C$, $Q_s'(u\to 1)\to \infty$, while for $k<k_C$, $Q_s'(u\to 1)\to -\infty$, we can find $k_C$ by demanding that at $k=k_C$, the first derivative at $u=1$  does not diverge. Using Eq.~(\ref{pdfSM}), we differentiate the logarithm of $Q_s(u)$ once and evaluate the result in the vicinity of $u=1$. In the limit of $\alpha\ll 1$, the result is
\begin{equation}
	\frac{d\ln Q_s(u)}{du}=\frac{Q_s'(u)}{Q_s(u)}\simeq  \frac{2/[k(1+\delta)]-1}{u-1}+\frac{\delta-1}{\alpha(\delta+1)}+{\cal O}(u-1).
\end{equation}
As stated, at $k=k_C$ the first (diverging) term has to vanish. Therefore, we find
\begin{equation}
	k_C = \frac{2}{1+\delta}.
\end{equation}
This result consistently reduces to $k_C=1$, at $\delta\to 1$. Note, that the value of $k$ at which the PDF flips at $u=-1$ is lower than $k_C$ and is obtained at $k=2\delta/(1+\delta)<k_C$. Only for values of $k$ lower than this value, the PDF actually becomes unimodal, and the mode is given by Eq.~(\ref{mode}), which is indeed valid as long as $u_*>-1$ or $k<2\delta/(1+\delta)$.

\bibliographystyle{apsrev4-1}
\bibliography{literature}

\end{document}